\documentclass[aps,prl,twocolumn,superscriptaddress]{revtex4}

\usepackage{hyperref}
\usepackage{amsmath,amssymb,amsfonts}
\usepackage{color}
\usepackage{graphicx}
\usepackage{enumerate} 
\usepackage{colordvi} 
\usepackage{bm}

\newcommand{\ihgpc}{\, h\, {\rm Gpc}^{-1}}

\newcommand{\vx}{\mathbf{x}}

\newcommand{\vk}{\mathbf{k}}

\newcommand{\vecr}{\mathbf{r}}

\newcommand{\vv}{\mathbf{v}}

\newcommand{\wt}{\widetilde}

\newcommand{\himpc}{{\hbox {$~h^{-1}$}{\rm ~Mpc}}}

\newcommand{\be}{\begin{equation}}
\newcommand{\ee}{\end{equation}}
\newcommand{\bey}{\begin{eqnarray}}
\newcommand{\eey}{\end{eqnarray}}

\newcommand{\paper}{paper }

\begin{document}
\title{
Intrinsic Alignments and Splashback Radius of Dark Matter Halos \\
from Cosmic Density and Velocity Fields
}
\author{Teppei Okumura}\email{tokumura@asiaa.sinica.edu.tw}
\affiliation{Institute of Astronomy and Astrophysics, Academia Sinica, P. O. Box 23-141, Taipei 10617, Taiwan}

\author{Takahiro Nishimichi}
\affiliation{Kavli Institute for the Physics and Mathematics of the Universe (WPI), UTIAS, The University of Tokyo, Kashiwa, Chiba 277-8583, Japan}
\affiliation{CREST, JST, 4-1-8 Honcho, Kawaguchi, Saitama, 332-0012, Japan}

\author{Keiichi Umetsu}
\affiliation{Institute of Astronomy and Astrophysics, Academia Sinica, P. O. Box 23-141, Taipei 10617, Taiwan}

\author{Ken Osato}
\affiliation{Department of Physics, University of Tokyo, 7-3-1 Hongo, Bunkyo-ku, Tokyo 113-0033 Japan}

\date{\today}

\begin{abstract}
We investigate the effects of intrinsic alignments (IA) of dark-matter halo shapes on cosmic density and velocity fields from cluster to cosmic scales beyond $100\himpc$. Besides the density correlation function binned by the halo orientation angle which was used in the literature, we introduce, for the first time, the corresponding two velocity statistics, the angle-binned pairwise infall momentum and momentum correlation function. Using large-volume, high-resolution $N$-body simulations, we measure the alignment statistics of density and velocity, both in real and redshift space. We find that the alignment signal is not amplified by redshift-space distortions at linear scales. Behaviors of IA in the velocity statistics are similar to those in the density statistics, except that the halo orientations are aligned with the velocity field up to a scale larger than those with the density field, $x>100\himpc$. On halo scales, $x\sim R_\mathrm{200m} \sim 1\himpc$, we detect a sharp steepening in the momentum correlation associated with the physical halo boundary, or the splashback feature, which is found more prominent than in the density correlation. Our results indicate that observations of IA with the velocity field can provide additional information on cosmological models from large scales and on physical sizes of halos from small scales.  
\end{abstract}
\pacs{98.80.-k}
\keywords{cosmology, large-scale structure} 
\maketitle

\flushbottom
{\it Introduction.}--- Intrinsic alignments (IA) of galaxy images in principle contain useful information to probe galaxy formation processes. Detailed studies of IA are also important on large scales since it can act as contamination when the shape information is used in weak lensing surveys for precision cosmology. The effects of this contamination have been actively studied both theoretically \cite{Catelan:2001,Hirata:2004} and observationally \cite{Mandelbaum:2006,Okumura:2009,Li:2013,Singh:2015} (See, e.g., \cite{Schafer:2009,Troxel:2015} for reviews). More interestingly, this IA effect is known to depend strongly on the mass of host dark-matter halos \cite{Jing:2002}. Thanks to robust and efficient cluster-finding algorithms recently developed based on the observed galaxy distribution \cite{Rykoff:2014,Oguri:2014}, strong alignment signals have been detected up to scales larger than $100\himpc$ \cite{van-Uitert:2017,Huang:2017}. Galaxy clusters are thus ideal objects to address a fundamental question of up to what scales luminous objects are aligned with the matter distribution in the large-scale structure of the universe. Moreover, clusters enable us to investigate intra-halo structure by cross-correlating with galaxies in and around the clusters. Here, of particular interest is the identification and characterization of physical boundaries of halos by studying splashback features associated with the outermost caustic in accreting halos \cite{Diemer:2014,More:2016,Umetsu:2017}. The goal of this study is to simultaneously investigate the splashback features and IA of clusters.   

In past work, alignments of the major axes of galaxies relative to the overdensity field have been studied when IA were investigated.  However, it is fundamentally important to consider the IA relative to the cosmic velocity field with various reasons. Now the velocity field at cosmic scales can be measured through several ways, such as peculiar velocity \cite{Strauss:1995} and kinematic Sunyaev-Zel'dovich (kSZ) \cite{Hand:2012} surveys. Since the velocity field in Fourier space follows $\vv_\vk \propto (i\vk/k^2) \delta_\vk$ in linear theory, one expects that the velocity correlation signal is amplified compared to the density counterpart on large scales. This trend has been seen in the measured pairwise velocity power spectrum in kSZ surveys \cite{De-Bernardis:2017,Sugiyama:2017b}. In this \paper we introduce new statistics, namely the alignment velocity statistics. We formulate the two alignment velocity statistics, and using the high-resolution $N$-body simulations, we test the statistics over a broad range of scales, from a 1-halo regime to the scale beyond $100\himpc$.  

{\it Alignment correlation function.}--- The statistic commonly used to quantify IA in weak-lensing surveys is the correlation between the gravitational shear and intrinsic ellipticity (GI) \cite{Hirata:2004}, sampled by two fields $A$ and $B$, respectively \footnote{Here we consider a general case where the two fields are different, e.g., $A$ and $B$ are the fields traced by galaxies and galaxy clusters, respectively, but the expression for the auto correlation can be obtained by setting $A=B$.}, 
$\xi_{A+}(\vx)=\left\langle\delta_A(\vecr_1)\gamma^I(\vecr_2)\right\rangle$, where $\vx={\bf r}_2-{\bf r}_1$, $\gamma^I(\vecr)=q\cos(2\theta)$,   
$\theta$ is the angle between the elongated orientation of the matter field, traced by the major axes of halo or galaxy shapes, and $\vx$ projected onto the sky, and $q$ is a weight factor of the major-to-minor-axis ratio.

In this work, we consider an alternative quantity, namely the angle-binned or alignment correlation function \cite{Paz:2008,Faltenbacher:2009}
\footnote{We use the term ``angle'' to indicate the projected angle $\theta$ instead of the direction with regard to the line of sight.}. 
It is defined as an extension of the conventional two-point correlation function of the fields $A$ and $B$, by taking account of the orientation of the field $B$, $\xi_{AB}(\vx,\theta)=\left\langle \delta_A({\bf r}_1)\delta_B({\bf r}_2,\theta)\right\rangle$. The conventional correlation function, 
$\xi_{AB}(\vx)=\left\langle \delta_A({\bf r}_1)\delta_B({\bf r}_2)\right\rangle$, 
can be obtained by integrating over $\theta$,
$\xi_{AB}({\bf x})=(2/\pi)\int^{\pi/2}_{0} d\theta \xi_{AB}({\bf x},\theta)$. 
Since the distances to objects are measured through redshift in galaxy surveys, the correlation function is affected by their velocities and becomes anisotropic, known as redshift-space distortions (RSD) \cite{Kaiser:1987}. Although RSD produce multipole moments, such as the quadrupole and hexadecapole, in this \paper we consider only the monopole,
$\xi_{AB,0}(x,\theta)=\int^1_0 d\mu \xi_{AB}(\vx,\theta)$, where $\mu$
is the direction cosign between the vector $\vx$ and the line of sight, $\mu=\hat{\vx}\cdot\hat{\vecr}$, where the hat denotes a unit vector. Throughout this paper we assume the distant observer approximation, so that $\hat{\vecr}_1=\hat{\vecr}_2\equiv\hat{\vecr}$. Hereafter we omit the subscript $0$ which represents the monopole ($\ell = 0$). The GI function corresponds to the dipole moment of the alignment correlation function, $\wt{\xi}_{A+}(\vx)=(2/\pi)\int^{\pi/2}_{0} d\theta \cos(2\theta)\xi_{AB}(\vx,\theta)$, where $\wt{\xi}_{A +}$ is the same as $\xi_{A +}$ but with $q$ fixed to $q=1$ \cite{Okumura:2009a,Faltenbacher:2009}.

{\it Alignment velocity statistics.}--- In analogy to the density statistic, in this \paper we consider two alignment statistics of halos/galaxies relative to the cosmic velocity field. One is the pairwise mean momentum \cite{Fisher:1995,Okumura:2014}:
$p_{AB}(\vx)=\left\langle \left[1+\delta_A ({\bf r}_1)\right] \left[1+\delta_B ({\bf r}_2)\right]\left[v_{A}^z ({\bf r}_1)-v_{B}^z ({\bf r}_2)\right]\right\rangle$,
where $v_A^z(\vecr)$ is the radial component of the peculiar velocity of sample $A$, $v_A^z(\vecr)=\vv_A(\vecr)\cdot \hat{\vecr}$. Another statistics is the momentum correlation function \cite{Gorski:1988}:
$\psi_{AB}(\vx)=\left\langle \left[1+\delta_A ({\bf r}_1)\right]\left[1+\delta_B ({\bf r}_2)\right]
v_{A}^z ({\bf r}_1)v_{B}^z ({\bf r}_2)\right\rangle$. Note that in observations such as peculiar velocity and kSZ surveys, the velocity field is also sampled in redshift space, thus affected by RSD, as formulated by \cite{Okumura:2014,Sugiyama:2016}.

We define the angle-binned pairwise infall momentum, $p_{AB}(\vx,\theta)$, and momentum correlation function, $\psi_{AB}(\vx,\theta)$, by replacing $\delta_B(\vecr_2)$ in the above equations by $\delta_B(\vecr_2,\theta)$ 
\footnote{We also defined the the angle-binned density-momentum cross correlation,
$\left\langle \left[1+\delta_A ({\bf r}_1)\right]  \left[1+\delta_B ({\bf r}_2,\theta)\right] v_{A}^z ({\bf r}_1)\right\rangle $,
and angle-binned, pair-weighted velocity dispersion,
$\left\langle \left[1+\delta_A ({\bf r}_1)\right]\left[1+\delta_B ({\bf r}_2,\theta)\right]\left(v_{A}^z ({\bf r}_1)-v_{B}^z ({\bf r}_2)\right)^2\right\rangle$, 
and investigated their properties. The behaviors of IA in the former and latter statistics were, however, found 
similar to those in $p_{AB}$ and $\psi_{AB}$, respectively.}.
Similarly to the density correlation function, the conventional expressions for these velocity statistics, $p_{AB}(\vx)$ and $\psi_{AB}(\vx)$, can be obtained by averaging over $\theta$. These are the main statistics we will study in this \paper. We will focus only on the lowest-order moments of these statistics, namely the dipole of the pairwise momentum and the monopole of the momentum correlation function, $p_{AB}(x,\theta)=3\int^1_{0} d\mu \mu p_{AB}(\vx,\theta)$ and $\psi_{AB}(x,\theta)=\int^1_{0} d\mu \psi_{AB}(\vx,\theta)$, respectively. The effects of IA on the higher-order moments will be studied in our future work. 

{\it $N$-body simulations.}---
In order to study the alignment statistics, we use a series of large and high-resolution $N$-body simulations of the $\Lambda$CDM cosmology seeded with Gaussian initial conditions. These are performed as a part of the \texttt{dark emulator} project \cite{Nishimichi:2017}. We adopt the cosmological parameters of $\Omega_m=1-\Omega_\Lambda = 0.315$, $\Omega_b=0.0492$, $h=0.673$, $n_s=0.965$, and $\sigma_8=0.8309$. We employ $n_p=2048^3$ particles of mass $m_p= 8.15875\times 10^{10}M_\odot / h$ in a cubic box of side $L_{\rm box} = 2\ihgpc$. We use 4 realizations in total, and analyze the snapshots at $z=0306$, which will be quoted as $z=0.3$ in the following.

Halos and subhalos are identified using phase-space information of matter particles, the {\it Rockstar} algorithm \cite{Behroozi:2013}. The velocity of the (sub)halo is determined by the average particle velocity within the innermost 10\% of the (sub)halo radius.  We use the standard definition for the halo radius and mass of $M_h \equiv M_{\rm \Delta m} = M(<R_{\rm\Delta m}) = (4\pi/3)\Delta{m} \rho_{\rm m}(z) R_{\rm\Delta m}^3$, with $\rho_{\rm m}$ the mean mass density of the universe at given redshift $z$, and we adopt $\Delta\mathrm{m} = 200$. To study cluster-scale halos, we select halos with $M_{h} \geq 10^{14}h^{-1}M_\odot$, which roughly corresponds to the typical threshold of the richness parameters used by the cluster-finding algorithms in the literature. We will mainly present the analysis of IA of the cluster shapes relative to the density field traced by galaxies. To this end, we create mock galaxy catalogs using a halo occupation distribution (HOD) model \cite{Zheng:2005} applied for the LOWZ galaxy sample of the SDSS-III Baryon Oscillation Spectroscopic Survey obtained by \cite{Parejko:2013}. We populate halos with galaxies according to the best-fitting HOD $N(M_h)$. For halos that contain satellite galaxies, we randomly draw $N(M_h)-1$ member subhalos to mimic the positions and velocities of the satellites (see \cite{Nishimichi:2014,Okumura:2017} for alternative methods). We assume halos to have triaxial shapes and estimate the orientations of their major axes using the second moments of the mass distribution projected onto the celestial plane. Table \ref{tab:halo} summarizes properties of our mock samples.

\begin{table}[bt!]
\caption{Properties of mock halo/subhalo samples at $z=0.3$. $f_{sat}$ is the number fraction of satellites, $M_\mathrm{min}$ and $\overline{M}$ are the minimum and average halo mass in units of $10^{12}h^{-1}M_\odot$, respectively, and 
$\overline{n}$ is the number density in units of $h^3{\rm Mpc}^{-3}$.}
\begin{center}
\begin{tabular}{|c|ccccc|}
\hline
halo types & $f_{sat}$ & $M_{\rm min}$ &   $\overline{n}$ &$b$ & $\overline{M}$ \\ 
\noalign{\hrule height 1pt}
Clusters (halos)  & 0 &100 & $ 2.05 \times 10^{-5}$ & $3.05$ & 188\\
Galaxies (subhalos)  & 0.137 & 1.63 &   $5.27\times 10^{-4}$ &$1.69$ & 25.2 \\ 
\hline
\end{tabular}
\end{center}
\label{tab:halo}
\end{table}

{\it Estimators.}--- 
In observational studies of IA in weak-lensing surveys, one usually considers a correlation function projected along the line of sight. Thus, the contributions of RSD are largely cancelled out, while the IA signal is also smeared out to some extent by projection effects. Since we are interested in the IA signal itself, we present the alignment density and velocity correlation statistics in both real space and redshift space. 

The alignment correlation function of the density ($A$) and shape ($B$) samples can be measured by
$\xi_{AB}(x,\theta)=\left\langle D_AD_B\right\rangle/\left\langle R_AR_B\right\rangle-1$, where 
$\left\langle D_AD_B\right\rangle(x,\theta)$ and $\left\langle R_AR_B\right\rangle(x,\theta)$ are respectively the normalized pair counts of the data and their randoms as functions of separation $x$ and angle $\theta$. The momentum correlation function can be measured by an estimator:
$\psi_{AB}(x,\theta)=\left\langle V_{A}^zV_{B}^z\right\rangle/\left\langle R_AR_B\right\rangle$,
where 
$\left\langle V_{A}^zV_{B}^z\right\rangle(x,\theta)$ is the normalized pair count of Samples $A$ and $B$ weighted by the products of the radial components of their velocities as functions of $x$ and $\theta$. We adopt an estimator for the pairwise momentum dipole \cite{Ferreira:1999,Okumura:2014}, 
$p_{AB}(x,\theta)=\left\langle(V_{A}^zD_B-D_AV_{B}^z)\hat{\vecr}\cdot\hat{\vx}\right\rangle/\left\langle R_AR_B(\hat{\vecr}\cdot\hat{\vx})^2\right\rangle$.
In this analysis, the shape information is taken from clusters, so that sample $B$ is always a cluster, $B=\{c\}$, while otherwise stated, the density field $A$ is traced by galaxies, $A=\{g\}$. In the following analysis, we measure these statistics from each of the four realizations and present their means and standard errors from the scatters.

\begin{figure}[bt]
\includegraphics[width=0.49\textwidth,angle=0,clip]{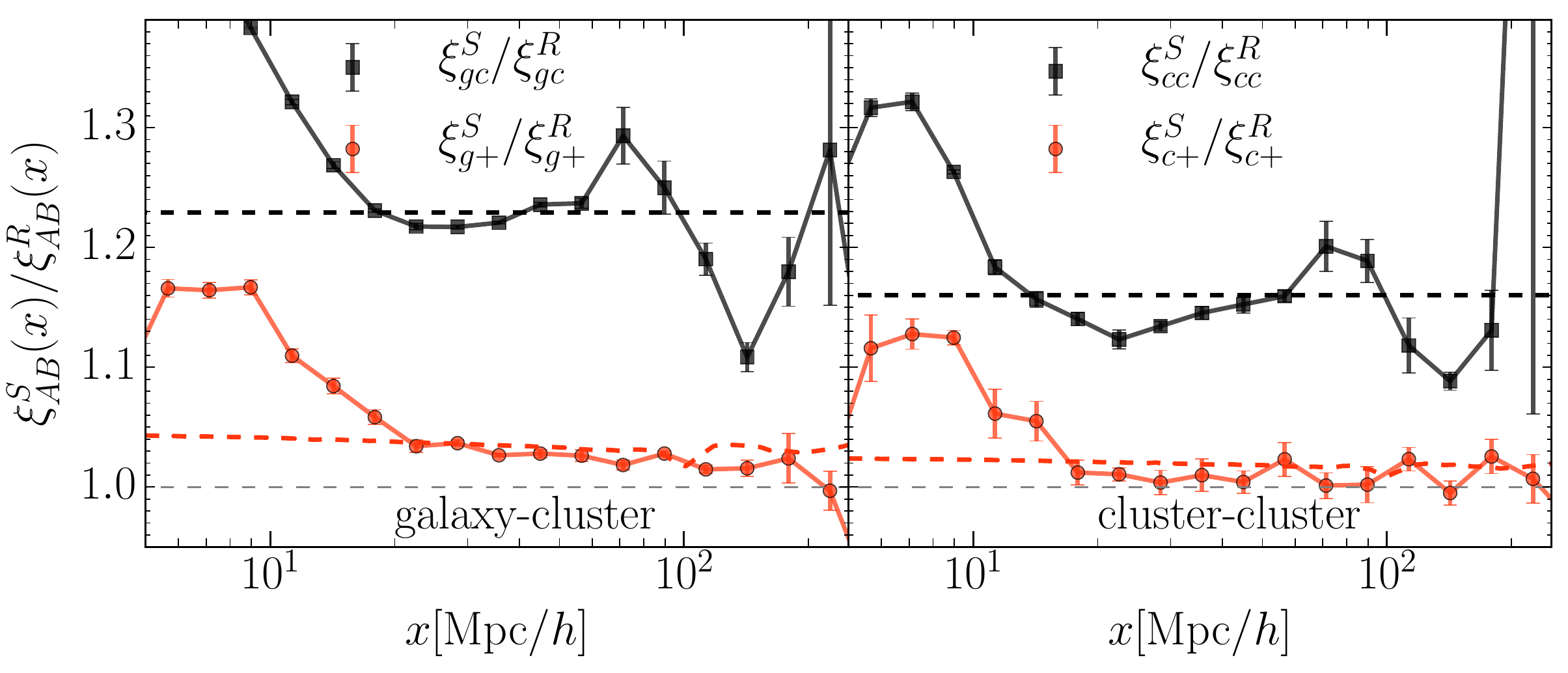}
\caption{Ratio of GI correlation functions in redshift and real space for galaxy-cluster (left) and cluster-cluster (right), shown as the red points. The corresponding ratios of the density correlations are shown as the black points. The black and red lines are the linear Kaiser predictions.
}
\label{fig:gi_rs_2gpc_z014}
\end{figure}

{\it Numerical analysis.}--- We start by presenting the GI correlation functions in real and redshift space because the results can be compared to simple analytical predictions \cite{Singh:2015}. The left panel of Fig.~\ref{fig:gi_rs_2gpc_z014} shows the ratio of the GI correlation functions in redshift and real space measured using the estimator of \cite{Mandelbaum:2006}. The right panel shows the result where both the shape and density fields are from the clusters. For comparison, we show the corresponding ratios of the density correlation functions in redshift and real space. The horizontal black line is the linear prediction of Kaiser formula \cite{Kaiser:1987}, $\xi_{AB}^S/\xi_{AB}^R=1+(\beta_A+\beta_B)/3+\beta_A\beta_B/5$, where $AB=\{\rm gc,cc\}$, $\beta_A=f/b_A$, $f=\ln{\delta}/\ln{a}$ with $a$ the scale factor, and the superscripts $S$ and $R$ denote the statistics defined in redshift and real space, respectively. The linear bias parameter of Sample $A$ is determined from $b_A(x)=(\xi_{AA}^R/\xi^R_m)^{1/2}$ on $20<x<80 \,[\himpc]$ where $\xi^R_m$ is the matter correlation function and here we simply compute it using linear theory based on the {\it CAMB} code \cite{Lewis:2000}. The values of the bias are shown in Table \ref{tab:halo}. The Kaiser prediction gives a reasonable ratio around $1.2$ for the density correlation on sufficiently large scales up to $x\sim 125\himpc$, where the correlation crosses zero. The enhancement of the 3D alignment statistics by RSD is, by contrast, only a few percent for the galaxy-cluster GI, and further suppressed for cluster-cluster GI by the factor $f/b$, as shown as the red lines. The smaller ratio for the alignment statistics is because of the more complicated oscillatory kernel in the integral. In the following analysis we will show the alignment correlations of density and velocity measured both in real and redshift space, although the alignment signals are expected to be similar according to this result. 

\begin{figure}[bt]
\includegraphics[width=0.49\textwidth,angle=0,clip]{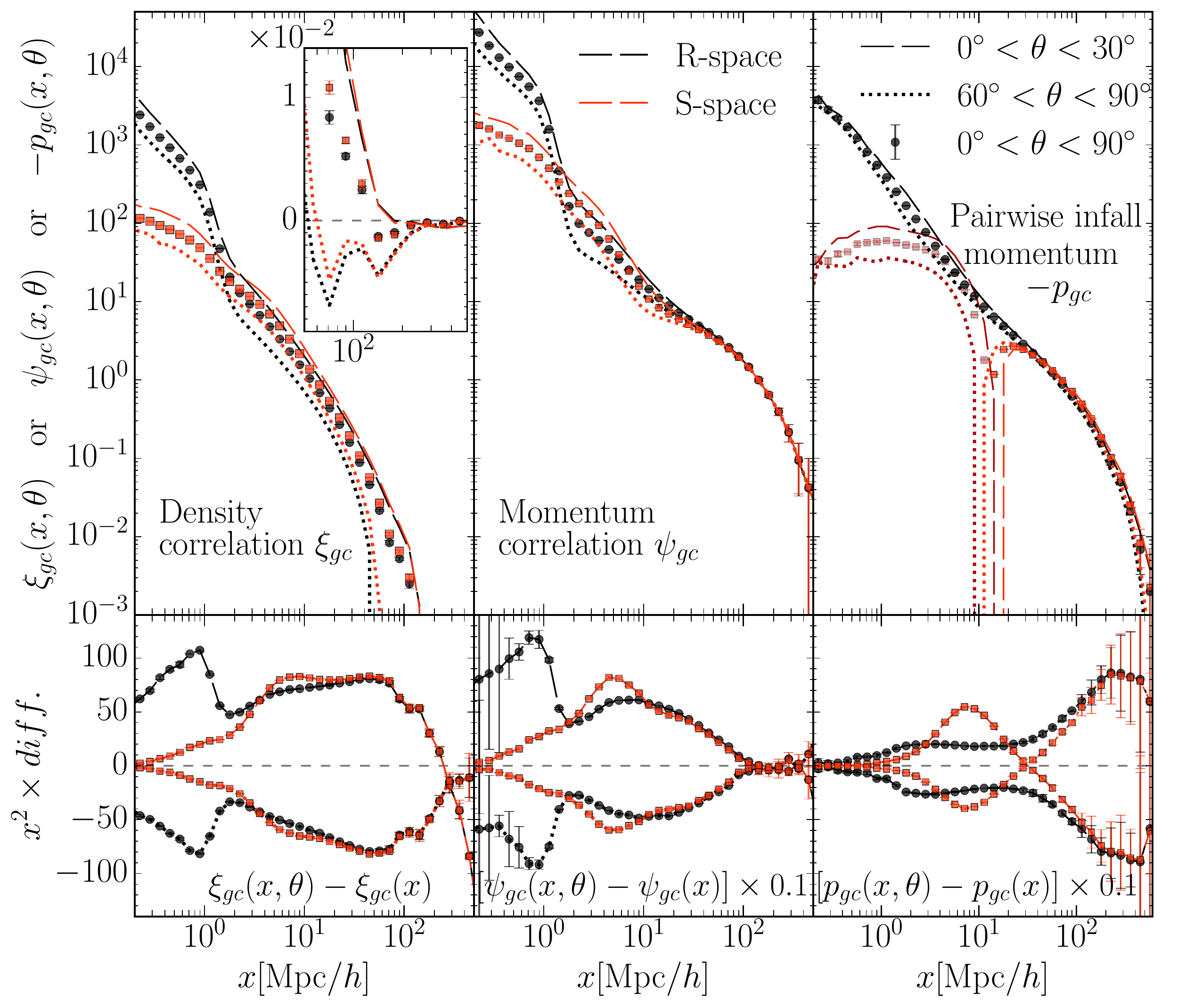}
\caption{(Left) 
Angle-binned density correlation function of galaxy density and cluster orientation in real (black) and redshift (red) space. In the top panel the dashed and dotted lines respectively show the correlations parallel and perpendicular to the major axes of the clusters $\xi_{gc}(x,\theta)$ while the points with error bars the conventional correlation functions, and the bottom panel is their difference, $x^2[\xi_{gc}(x,\theta)-\xi_{gc}(x)]$. The inset provides the zoomed view on large scales. 
(Center) Angle-binned momentum correlation functions, $\psi_{gc}(x,\theta)$, which have a unit of $[\himpc]^2$. In the bottom panel the amplitude is multiplied by $0.1$.
(Right) Angle-binned pairwise mean infall momentum. Since the statistics has negative values at linear scales, we show $-p_{gc}$ at the top panel, and the points and lines colored darker-red are its negative values in redshift space, $-(-p_{gc}^S)=p_{gc}^S$.
}
\label{fig:xiab}
\end{figure}

In the top-left panel of Fig.~\ref{fig:xiab}, we show the cluster-galaxy density cross-correlation function binned in $\theta$. Though not presented here, the trend is basically the same for the cluster auto correlation. As expected from Fig.~\ref{fig:gi_rs_2gpc_z014}, the alignment signals are almost the same on large scales in real and redshift space unlike the density correlation. This signal has been measured both in simulations and observations \cite{Schneider:2012,Li:2013}. 

\begin{figure}[bt]
\includegraphics[width=0.49\textwidth,angle=0,clip]{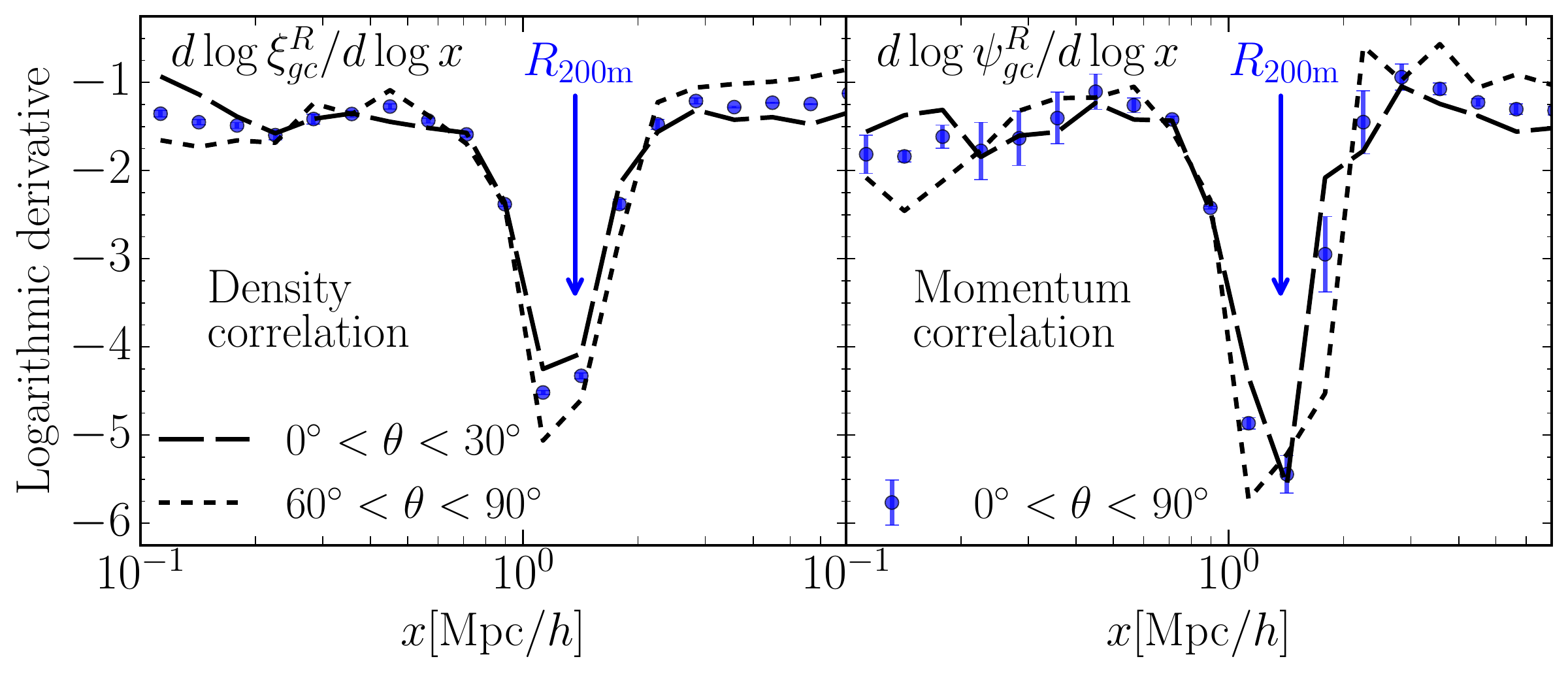}
\caption{(Left) Logarithmic slope of the angle-binned density correlation of galaxy density and cluster orientation in real space. The solid and dotted lines, respectively, are the correlation parallel and perpendicular to the major axes of the clusters, while the the blue points with error bars are for the conventional correlation function. The location of $R_\mathrm{200m}$, which coincides well with $R_{sp}$, is denoted by the arrow.
(Right) Same as the left panel but for the momentum correlation function. 
}
\label{fig:xi_splash}
\end{figure}

On the scale of $x\sim 1\himpc$, an abrupt change in the slope of the real-space correlation can be seen, whereas it is washed out in redshift space by the $\sim 10\himpc$ line-of-sight displacement of galaxies due to RSD. There have been intensive efforts to understand and characterize the physical boundary of halos in terms of the steepest point of the density gradient. This steepest-slope location is a signature of the splashback radius $R_\mathrm{sp}$ \cite{Diemer:2014,More:2015a} (See \cite{Okumura:2015} for an alternative approach to define the boundary using the satellite distribution). The left panel of Fig. \ref{fig:xi_splash} presents the logarithmic slope of the real-space alignment correlation, $\gamma = d\log{\xi_{gc}^R}/d\log{x}$. The steepest slope of our conventional correlation function reaches $\gamma \simeq -4.5$, significantly steeper than the asymptotic outer slope ($-3$) of the standard Navarro--Frenk--White profile \cite{Navarro:1996}. This is consistent with the characteristic splashback properties in $\Lambda$CDM found by \cite{Diemer:2014}. The location of $R_\mathrm{200m}$ in our sample, which coincides well with the respective steepest gradient point, is denoted by the arrow in the figure. The mean $R_\mathrm{sp}$--$R_\mathrm{200m}$ relation of \cite{Diemer:2017a} predicts $R_\mathrm{sp}=1.001R_\mathrm{200m} \simeq  1.37 \himpc$ for our cluster sample with $\overline{M}_\mathrm{200m}=1.88\times 10^{14} h^{-1}M_\odot$, so that it precisely matches the splashback scales found here. Intriguingly, the slopes are shallower and steeper, respectively, for the alignment correlation parallel and perpendicular to the major axes of the clusters. This can be qualitatively interpreted by the fact that the halo size varies more significantly along the direction of the major axis (i.e., smaller $\theta$), leading to a less prominent boundary. It is also interesting to note that the IA signal is further enhanced within the splashback shell, as seen in the left bottom panel of Fig.~\ref{fig:xiab}, consistent with the halo model of IA \cite{Schneider:2010}.

The next quantity to consider is the angle-binned momentum correlation function. Its numerical results are shown in the upper-middle panel of Fig.~\ref{fig:xiab}. On large scales where linear theory holds, the correlation in redshift space is equivalent to that in real space \cite{Okumura:2014}. As seen in the lower-middle panel, the IA signal in the momentum correlation vanishes at $x\sim 100\himpc$, on a scale smaller than that of the density correlation. We see a steepening feature of the real-space momentum correlation at $R_\mathrm{sp}\simeq R_\mathrm{200m}$, very similar to that of the density correlation. To see the feature in more detail, we show the logarithmic derivative, $\gamma_p=d\log{\psi_{gc}^R}/d\log{x}$, in the right panel of Fig.~\ref{fig:xi_splash}. It is interesting to note that the slope approaches $\gamma_p \to -5.5$, steeper than the boundary slope determined by the density correlation. As discussed by \cite{Lapi:2009} in the context of $\Lambda$CDM structure formation, the orbital velocity anisotropy is tightly coupled with the logarithmic density slope around halos and thus expected to be sensitive to the location of the halo edge \cite{Diemer:2014}, which physically and sharply separates the multi-stream intra-halo region from the outer infall region (see also \cite{Faltenbacher:2010}). Besides, the splashback feature and the enhancement of IA at this scale are found more prominent for the momentum correlation than for the density correlation. However, since the momentum correlation is noisier than the density correlation, we cannot see a clear dependence of the splashback feature on IA in our simulations.

Finally, we turn to another velocity statistic, the pairwise mean infall momentum. The upper-right panel of Fig.~\ref{fig:xiab} shows the angle-binned pairwise mean momentum. As expected, we do not see a splashback feature from this statistic. On small scales, the sign of the redshift-space pairwise momentum changes from negative to positive due to the non-linear velocity dispersion \cite{Okumura:2014,Sugiyama:2016}. On the other hand, objects approach each other on average, so that the correlation becomes negative on large scales. Note that in the lower-right panel the alignment signal in $p_{gc}$ persists beyond $x\sim 200\himpc$.  

{\it Outlook.}--- The presented analysis of IA with the velocity field has several important applications. Recently, an idea of testing inflation models using IA on large scales has been proposed by \cite{Schmidt:2015}. The angle-binned pairwise momentum, $p_{AB}(x,\theta)$, or its ratio to the density correlation, may serve as a powerful tool to probe inflation. Future kSZ surveys enable us to measure the large-scale velocity field and constrain cosmological models using the higher-order velocity moments \cite{Sugiyama:2017}. Detailed studies of the velocity IA effect based on (non)linear alignment and halo models \cite{Hirata:2004,Schneider:2010,Blazek:2011}, as well as large-volume $N$-body simulations, will be presented in our future work. A possible contaminant of utilizing the kSZ is the effect of the optical depth. However, a promising method of measuring it has recently been proposed based on a semi-analytic technique calibrated with X-ray observations \cite{Flender:2017}. Note that to compare simulation results of IA to observations, shapes of brightest cluster galaxies (BCG) are often used as a proxy for the cluster shapes. In this case, misalignment between the major axes of BCG and their host halos needs to be taken into account \cite{Okumura:2009}. Although the misalignment for such massive halos is known to be small \cite{Despali:2017}, it needs to be studied carefully.  

The splashback radius $R_\mathrm{sp}$ has been detected for the first time in the momentum correlation function from simulations. The feature is even sharper than that in the density correlation, because $R_\mathrm{sp}$ separates distinct infall and multi-stream regions of collisionless CDM. High-resolution kSZ observations can be used to test this new prediction on the nature of DM in comparison to the splashback feature observed in the density correlation \cite{More:2016}, as well as the dependence of cluster orientations on $R_\mathrm{sp}$.

\begin{acknowledgements} We thank Benedikt Diemer for useful comments. This work was in part supported by MEXT Grant-in-Aid for Scientific Research on Innovative Areas (No.~15H05887, 15H05893, 15K21733, and 15H05892). T.N. acknowledges financial support from JSPS KAKENHI Grant Number~17K14273 and JST CREST Grant Number~JPMJCR1414. K.U. acknowledges support from the Ministry of Science and Technology of Taiwan under the grant MOST 103-2112-M-001-030-MY3. \end{acknowledgements}

\bibliography{ms.bbl}
 \end{document}